# On the construction of a psychologically based, general theory of observation: an introduction

By Göte Nyman

Key words: Observer theory. Perception. Observer physics.

Correspondence: Göte Nyman, Östermalmintie 49, 02580 Siuntio, Finland

gote.nyman@helsinki.fi

## Introduction

The progress of modern physics including the theory of relativity and quantum mechanics has been accompanied by problems, argumentations, and thought experiments that directly deal with the act of human observation and measurement. However, it may come as a surprise that there is no generally accepted and explicit theory of the general observer in physics. Often it has been taken as given that an overall model of a human observer is sufficient. This paper offers a preliminary introduction to the perception-psychological arguments that highlight the importance of developing the observer theory that joins physics with perceptual psychology.

Historically, the human perceptual capabilities are grounded in the intertwined evolution of the human physiology and the cultural history of mankind. Only an idealized perceptual system (cf. Geisler 1989, 153) can make observations with optimal efficiency. Natural perception systems just come close to this if they have achieved ecologically feasible computational capabilities to match the observation needs and the nature of the available information. At the single cell level, receptor physics determine the absolute sensory limits of perception and in one specific case it has even been possible to (indirectly) relate the sensitivity of an individual receptor to quantum level processes, i.e. photon capture by an isolated retinal receptor cell in situ (Baylor et al. 1979, 613). However, for a genuine visual sensation to occur, more photons distributed over several rod receptors are required for an optimal 1 millisecond test flash to cause a cellular response signal that passes through the neural system and evokes an elementary visual sensation of light (Hecht et al. 1942, 196). Furthermore, thermal rhodopsin isomerization noise has been shown to affect the absolute visual sensitivity in studies that compared the sensory thresholds of animal species at different body temperatures. A clear correlation was been found for their absolute visual thresholds and the estimated thermal isomerization rate in the retina (Aho et al. 1988, 348).

## Errare humanum est: perception is estimation

It is not known what were the measurement practices before any measurement tools and standards were in use. Without standards, the perceptual classification of object properties is problematic because the human senses are very inaccurate in making physical estimations of e.g. color (reflectance), distance, or weight but also of time. Hence, the first pre-scientific concepts and symbols in communicating about the object world must have relied on the simplest and most reliable perceptions and communication symbols that could be shared by a community.

For example, color names emerged in almost every language (Kay, Maffi 1999, 743) and in a systematic order so that it became possible to communicate on object color properties by using basic color names, first using symbols for black and white and later for other colors as well. Apparently a hierarchy evolved in languages so that, for example, if a language had a term for red, it also had a term for black and white but the origin of this hierarchy is still under discussion (Berlin, Kay 1969) (Loreto et al. 2012, 1).

Whatever the way of communicating about the world when no measurement standards were available, the symbols used were inherently perception-related. Color names, for example, in the early languages were undoubtedly useful although relatively coarse and biased by many contextual factors such as lighting, shadows, surface reflections and contrasts. Based on the findings from language studies we may well hypothesize that the black-white (dark-bright) dimension of color classification was among the first pre-scientific and shared concepts that could be used to describe natural object properties in a standard-like manner. During this early evolution the subjective versions of the first physical concepts evolved. However, without solid references large perceptual attribute estimation errors occurred that could be 20% -50% and sometimes even more in estimating object distance, as is still known to happen in the modern contexts and even for experienced, professional observers such as pilots (cf. Foyle, Kaiser 1991, 314).

It is beyond the scope of the present paper to speculate about all other possible perceptual candidates that could have preceded the first pre-physical measurement standards. There are many of them and the concept of distance, for example, could have been based on the capacity of the 3D stereovision and scene analysis, walking distance or other motor performance metrics and the use of natural environmental references.

In everyday life and over the history of sciences, people have relied on the culturally shared practices for quantifying the world. As Kant (1781) put it *"Without community, each perception of an appearance in space is broken off from every other, and the chain of empirical representations—i.e. experience—would have to start all over again with each new object, with its immediate predecessor having not the least connection with it or being temporally related to it."* (translation by J.F. Bennett, Internet).

In other words, our observations and the measurements of the world make sense only if they can be shared within a (scientific) community. This sharing became possible when the early and pre-scientific communities adopted practical concepts and routines and started using them in situations that required an agreement on the amount and quality of objects and materials. In the Indus Valley (Baber 1996, 23) and Egypt the first measurement systems for length, time and weight (mass) were created between around 5$^{th}$ and 3$^{rd}$ millenium BC (cf. Iwata 2008, 2254) (Encyclopedia Britannica 2012). They paved the way to scientific measurements and observation practices.

What were the consequences of these cultural demands of better observer accuracy on the construction of the formal measurement systems and on the emerging physical theories? Was it just a matter of improving the observations with better measurement concepts and tools according to the code of the good scientific realist as Einstein pointed out: *"The belief in an external world independent of the perceiving subject is the basis of all natural science"* (Einstein 1954, 266)

A silly thought experiment

The way humans observe the world does not make sense to a frog community. Frogs have strongly derivative eyes that do not react to 'bugs' that don't move or show peculiar temporal characteristics (Lettvin et al. 1959). With a slight exaggeration and for the sake of argumentation we can say that a hungry frog will die in front of a delicious piece of food if the food does not move. Even if the frog had a human brain it would not help it to survive unless it had learned to use the neural output from its eye to the brain in an intelligent and creative way. But that would not be easy since its derivative eye is not spatially homogenous and it is not a linear system. What is lost early in the sensory-neural processing phase may not be possible to recover later. We can now ask what kind of physics would such a creature – a frog with frog eyes, but a human brain - develop when it could not have relied on a similar analysis of object position or size that we humans have learned to use? It would simply be blind to a measurement stick that does not move or flicker optimally. The problem becomes less silly if we forget the poor frog, and generalize the example to ask: How does an observer of a specific observer class perceive the world? How should we describe such an observer? How would the observer characteristics guide and constrain the empirically driven physical theories of the world that it can build?

In the classical Newtonian physics there is a clear-cut division between the object and the observer and the assumption that the observer does not have a direct impact on the object. The role of the observer is just to record whatever information is available for it to sense of the environment. If the observer does not have suitable sensory capacities it is doomed to remain blind to part of the world and only measurement tools that extend its sensory domain or spectrum (still remaining within it) can be of help. However, in quantum mechanics (some form of) entanglement between the object and the observer is unavoidable. The observer and the object together form a system that due to the process of observation will adopt a state that is dependent on both of the participating systems. Because of this the nature of the observer is critical to the outcome of the entanglement. Quite surprisingly and despite this observation-dependent nature of the quantum world, the physicists have generally assumed (often without explicit mentioning of it) that the exact evolutionary and adaptive nature of the human or animal perceptual system itself has little relevance to the theories of observation. To the best of my knowledge no explicit observer-perceiver model has been suggested for the analysis of quantum mechanical observations and interactions.

There is a well-known psychophysical observer theory that is aimed at formalizing and describing the observer as capable of performing intelligent and relevant perceptual inferences about the objective world, including quantum system contexts. It is a general psychophysical theory that is claimed to apply to every perceptual capacity. It offers a means to analyze the decision making potential of general psychophysical observers that are confronted with a perceptual problem in a world that has an objective description and solution and that can be described in terms of a set of propositions and possible observables related to that reality (Bennett et al. 1989, 231).

Bohr, Einstein, Heisenberg, and Schrödinger did have extensive discussion concerning the observation phenomenon. They were early to recognize that the act of observation remains a key problem in quantum mechanics and wondered what happens when a system (a human with a measurement equipment) observes and - by doing that - interferes with the quantum system. (Wimmel 1992)(cf. Lansman 2006, 212). Nevertheless, we don't know how the cultural-perceptual background of the observer and his knowledge community influence the interpretation of the observations and consequently, guide the formation of his physical theories.

The observer as a blind spot of modern physics

There seems to be a belief among the physicists (cf. Einstein 1954) that the cultural determinants of human observation are dwarfed by the power of mathematics, experimental empiria, and human imagination. The popular thought experiments by Einstein and Schrödinger, for example, did not analyze the observer's perceptual system characteristics in any depth.

Heisenberg was aware of the problem of defining the observer and considered the consequences of the cultural evolution of physical concepts (Heisenberg 1958, cf. Internet): *"Even if we realize that the meaning of a concept is never defined with absolute precision, some concepts form an integral part of scientific methods, since they represent for the time being the final result of the development of human thought in the past, even in a very remote past; they may even be inherited and are in any case the indispensable tools for doing scientific work in our time. In this sense they can be practically a priori. But further limitations of their applicability may be found in the future."*

He considered these inherited human concepts as part of a priori knowledge in science. However, these considerations neglected the possibility of a general theoretical framework that could point out the exact constraints that the human observer (or any observer) carries to the observation context.

In the special relativity theory (SRT) the observer is defined in terms of an inertial reference frame (Einstein 1905) within which he is assumed to preserve his 'classic' characteristics and to act as a reliable and relatively noiseless perceiver of time, mass, and distance. In this sense, he is considered as a real - perhaps even ideal - but at least not disturbingly biased observer located at a particular point in the space where he is trusted as a reliable observer of the incidences in his space-time system. If his senses are not sufficient then innovative measurement instruments like a clock can be of help. In his thought experiments Einstein assumed that the observer is a genuine classical physicist who does not suffer from the consequences of perceptual illusions or unconscious inferences (cf. Helmholtz 1867/1910). Perhaps he took it for granted that such natural perceptual phenomena would only complicate the thought experiments, but they would not alter their main messages – and he was right in the SRT context.

The observer-philosophical ideas from Kant were familiar to the fathers of modern physics who were puzzled by the role of a priori knowledge in observations. Despite this the observer-related theoretical forms of a priori knowledge were not explicitly formulated as it would have been relevant in order to fully describe the nature of the observations that involved an observer. A promising later candidate approach for explicit observer theory has been suggested by Caves et al. (2001, 1) in the form of Bayesian probability theory and based on the general argument that "quantum states are states of knowledge" (Fuchs, Peres 2000, 70) even though no explicit observer theory has been included there either.

Hence, it appears that the observer theories in physics have remained loosely human centered or human-specific, and one might even see them as somewhat speculative and detached from the empirically founded theories of perception. On the other hand there is an abundance of discussions under the term 'consciousness and quantum physics' or 'the brain and quantum physics' (cf. Penrose 1989) (Stapp 2007) that has continued up to the present time in cognitive sciences, philosophy and physics.

Quantum world of perception

Von Neumann suggested that for the wave function collapse to occur, a conscious mind of an observer is needed to receive information from the measurement (Von Neumann 1955). Heisenberg, however, thought that the transition from the 'possible' to the 'actual ' takes place at the moment of observation but without the contribution of the mind of the observer (Heisenberg 1958). He assumed that while observation (measurement) causes an interaction between the object and its observer the wave function collapse is not caused by the human mind as such. The observer was not explicitly formulated.

In Schrödinger's thought experiment the wave form collapse is thought to cause an apparent paradox where – without external observation – the isolated cat in the chamber enjoys the superposition principle and mysteriously remains potentially both alive and dead – if not observed. But again in this example, both the nature of the observation and the observer, to say nothing about the perceptual abilities of the cat itself, remain undefined and no explicit theory of the observer is included in the analysis. Bohr did consider the whole Schrödinger box as the *observer proper* that had 'seen' or been entangled with the destiny of the poor cat already before the imagined experimenter opened the box. This was a practical way to define 'observation' as a general relationship between two material entities: the cat and the box. However, no explicit relationship-theoretical formulation of the 'box as the observer' was offered. But clearly, if this assumption of the 'box as an observer' is accepted it implies the idea that any material entity can be an observer. If we then accept the self-evident idea that physics is a science that has been constructed by human observers then constructing a physics based on the characteristics of a material object such as a box is quite a puzzle. The giants of modern physics remained confused by the relationships between a real observer, its environment, and the observer's mind. This is not surprising considering the nature of this vastly complex problem field and the fragmented approaches in psychology and philosophy of these phenomena.

Hugh Everett's observer

A notable early exception in this rather fuzzy discourse on observer properties was Hugh Everett, who in his unpublished (handwritten) manuscript "Introduction of observers" (Everett 1955, Internet) characterized an idealized observer with a memory and also formalized it:

*"We wish now to make deductions about the appearance of phenomena on a subjective level, to observers which are treated within the theory. In order to accomplish this it is necessary to identify some properties of such an observer (states) with subjective knowledge (i.e. perceptions)."*

He continues:

*"It will suffice for our purpose to consider our observers to posses memories (i.e. parts of a relatively permanent nature, where states are in correspondence with past experience of the observer)."* He also described what he meant by a "good observation" that is interactive in nature. He did not suggest a detailed perceptual characteristic of the observer but clearly assumed that the observer's memory includes the basic perceptual characteristics of an observer.

Everett introduces the observer state function $\psi^o$[…A, B, C. …], where A, B, C ... represent the past experiences of the observer, in a temporally ordered sequence. Accordingly, his idea was then to treat the interaction of the observer with the object physical system, which process itself becomes a definition of an observation. Further on he introduces the objective requirement that in the observation process the eigenstate for the observed system remains unchanged and that the observer state change is unique to each system state. He offers a formal description to an observation, *"an observation upon a system that is not in an eigenstate of the observation"*, by combining the observer state $\psi^o$[ …] and the object system state to form a final combined state, which then becomes the observation proper (for the formal wave equation solution see the ms by Everett). Now, a question remains, what constitutes an observer state?

Recent findings in quantum mechanics have brought scientists closer to working with direct observer-object system characteristics when it has become possible to control the state of the observation process. The 2012 Nobel price in physics was awarded partly for the achievement in constructing a measurement (observation) set-up that entangles the object of measurement in a controllable way (Sayrin et al. 2011, 73). The distorting effects of the measurement (observer) system could be adjusted by creating a quantum control system that kept the observer effect as weak as possible. No explicit observer theory was, however, implied or mentioned there but it is possible to interpret the controlled entanglement process to be specific to and a result of the interaction with a certain type of an observer.

Constructing observer based physics: a thought experiment

Imagine a possibility to return to the scientific ground zero, to the point in time when no systematic physical measurement standards or tools existed. We can construct a theoretical observer $O_i(S_{j=1,N})$ with N hypothetical sensory functions $S_j$ that he uses to observe the world. Note that in the case of the human observer these functions are not meant to be identical with the sensory (physiological) systems that underlie them. The reason to this view is that at the moment there is no applicable theory available that would allow unique mapping of physiological processes on these elementary perceptual functions. It is not exactly known how many sensory functions there are in the human perceptual system as a whole. However, $S_j$ are the de facto perceptual functions that constitute the observer domain and observer capacities and which influence the construction of any tangible physical measurement systems by the community of observers $O_i$.

Accordingly, the original physical measures and concepts and consequently, the present theories in physics have been constructed based on these capabilities of the human (homo sapiens, hs) observer $O_{hs}$(l, m, t). The sensory functions for perceiving length (l), weight (m) and time (t) have made it possible to observe only certain types of natural event and artificial experiment and to develop physical theories with increasing accuracy in measuring the properties of objects that are within the reach of this observation domain and its extensions. The quantum mechanical studies and the constructed quantum physical laws are no exception to this since they include $S_j$ as their implicit ingredient in the way quantum mass, momentum, and position have been defined.

It is not reasonable to assume that for any community of observers $O_i$ the evolution of the physical theories would have automatically led to the same classical or quantum mechanical findings, equations, constants, and laws. Instead, we can imagine alternative hypothetical physics $P_i$ each of which is the outcome of the perceptions and actions of a specific observers $O_i$:

$O_i(S_j) \rightarrow P_i$

The present physics $P_{hs}$ reflects the characteristics of $O_{hs}$ We can now ask, is it reasonable to assume that other physics, outside the domain of $P_{hs}$ could be relevant and possible for us to know or even speculate about? Assuming that we are indeed interested in them how would $O_{hs}$ observe and measure the phenomena that are based on and predicted by these alternative physics $P_i$ where $i \neq hs$ ? A complete solution cannot be just the widening of the range or spectrum of $S_{hs}$ to extend the observer characteristics (sensory functions) within $P_{hs}$ – a step outside the domain of $S_{hs}$ is required.

Another silly thought experiment

Consider a bee that has visual receptors that are sensitive to ultraviolet (uv) light that humans cannot see. In order to observe uv, $O_{hs}$ has learned to use the wavelength concept and constructs devices that map the uv recordings (observations) on some of its perceptual system functions $S_{hs}$. Typically the observations take place in the visual domain and by visual indicators, but of course auditory, tactile or any other sensory domain could be used as well. However, in this process the bee physics $P_{bee}$ is not only mapped on $P_{hs}$ but instead it is assumed that there is a straightforward correspondence between $P_{bee}$ and $P_{hs}$ and that we can trust that the bee's observer characteristics $O_{bee}(S_{bee})$ are not relevant to our analysis. To put it simply, we rely on the human-centric way of defining the world and assume that all observers share the same world that has measurable properties. Accordingly, by assuming that the physics $P_{hs}$ is sufficient to apply in this context, we can take the bee as a member of the Kantian perceptual community, and we can discard ideas of other physics. This is not problematic if we accept the hypothesis that the bees live in a world that can be completely described by $P_{hs}$ and that nothing valuable is lost when the (now prevalent) mapping $S_{bee} \rightarrow S_{hs}$ is performed. We could then compute various sensory functions that the bee might use in its uv vision, map them on $S_{hs}$, and everything could make sense in $P_{hs}$.

But what if we want to understand the physics of the bee world proper? Bees have no explicit physics, but in a thought experiment we can assume that by their mere survival the bees must have developed an implicit theory of physics, $P_{bee}$ according to which their life and behavior is organized. However, if they could construct an explicit physics its ingredients would not be the same as ours. Of course, the problem of observer-based physics does not make any sense if we can assume that $P_{bee}$ is just a subset of $P_{hs}$. But the question now arises: is there something in the implicit $P_{bee}$ that our human physics $P_{hs}$ does not capture and how we could and should we know it? We don't know how and exactly why these two different physics have evolved but we can assume that the observer characteristics have been their drivers.

Of course it is possible to think that there is no meaningful $P_{bee}$ and that the world presents its kind faces to humans and that all relevant bee properties and behaviors can be mapped on $P_{hs}$. However, in quantum physics the situation is complex because there the basic assumption about the character of the world and the observation process itself are at test. Hence, repeating this thought experiment for a general observer, especially in the quantum realm the problem can become theoretically inspiring. Studying these alternative physics is a mathematical journey and adventure worthwhile taking.

How did physical measures emerge?

John Mollon starts the chapter on color vision and color blindness in his book by stating that "We are all color blind", referring to the fact that we have 'only' trichromacy of vision. It is well known that color blind people who cannot see a difference between red and green in controlled conditions, for example, can still discriminate between them as object properties on the basis of other naturally occurring visual features such as their perceived lightness or other contextual factors. In other words, they are able to compensate for their sensory disabilities by using other dimensions of their own perceptual space to locate the red and green objects (Mollon 1982, 165). But even in this case the idea of 'object color' is taken as a given and nobody doubts that that normal and color blind people share the same physics.

Human senses are indeed bad objective measurement instruments, but they have other benefits. Typically the Weber fraction in sensory magnitude discrimination tasks varies roughly between 2%-5%. There are extreme situations where sensory comparison has hyper accuracy like in the case of judging the alignment of two vertical lines (vernier acuity) where line displacement of only a few sec of arc can be discriminated (Westheimer, McKee 1977, 941). This is about 10 times better resolution than the optical Rayleigh resolution limit and the human visual acuity for traditional test targets such as Snellen letters, or sinusoidal gratings (Campbell, Robson 1968, 554). In stereovision a disparity of only a few seconds of arc cause perception of depth (Westheimer 1992, 205). Hence, it has been beneficial for human observers to design instruments that use the more accurate visual comparisons of object attributes, like the pointer location on a graphic scale, for example.

Overcoming the human sensory shortcomings

Numerous historical notes describe how the human perceptual limitations were compensated for by practical physical measures and standards that were used in trade, construction work, and everyday life. The core idea in these measurement systems was to utilize the best sensory (visual) functions, especially difference perception to allow accurate sensory comparisons. For example, direct estimation of object length without a comparison standard would easily introduce a measurement error of about 10% and even more. The strongest visual illusions, for example, can cause an error close to 50% in the estimation of line length. Hence, the Egyptian Royal Cubit (appr. 2700 B.C.) standard for conducting visual comparisons was based on the forearm length that was about 52.5 cm and probably made possible a measurement accuracy of about 0.2% (1 mm) at one measurement, depending on how the visual comparison was conducted.

Some object properties like liquid volume were difficult to measure and in the ancient China, for example, it was accomplished by filling standard wooden barrels with a specified amount of liquid. Then by hitting the to-be-filled barrel and the similar standard barrel containing the known amount of the liquid and comparing the sounds it was possible to decide when the liquid volumes were equal. In other words, the lack of a visual ability to judge the volumes had an auditory solution. In a sense, the physics needed for the measurement of volumes was not mature enough but it was possible to circumvent this by operating within a physiologically more accurate sensory domain. Quite similarly, water clocks were designed to help visualizing the passage of time and to quantify and divide it to different intervals (cf. Needham, Joseph et al. 1959)

One of the most compelling problems of measurement in modern times was the speed of light that was impossible to observe directly. Before Louis Fizeau's ingenious methods using fast rotating wheels (1849) the early experiments by Galilei to study it on the earth had failed due to the slowness of the visual system. The mechanical arrangements were not sensitive enough to allow the measurements of such huge speeds. By avoiding the sensory limitations and relying on the observation of celestial phenomena and benefitting from the long distances that light had to travel it had been possible for Roemer (1675) to arrive at a good approximation (cf. http://en.wikipedia.org/wiki/Speed_of_light#History).

The intimate consequences of the human body -centered measures like the Egyptian length standards were later seen in the classical architectural systems where they were introduced into the subjective world of aesthetics. The harmony system created by the Roman architect Vitruvius (approximately 70BC-15BC) and used as guideline in building temples consisted of a system of relationships based on the relative sizes of the human body parts: *"It is worthy of remark, that the measures necessarily used in all buildings and other works, are derived from the members of the human body, as the digit, the palm, the foot, the cubit, and that these form a perfect number, called by the Greeks τέλειος."* (Vitruvius, Internet). He did not only suggest a harmony system but also defined the navel as the center of the human body and the origin of the body-centered coordinate system. Interestingly, we have found brain cells in area 7 of the monkey cortex cells that were activated by movements on the skin that apparently had the reference point in the navel of the monkey (Leinonen et al. 1979, 303).

In summary, at least the following human-centered procedures have been used to overcome the human sensory limitations in conducting physical measurements.

*A. Sensory augmentation* by relying on a comparison against a standard within a sensory domain (vision): human body parts, measurement sticks and gauge pointers.

*B. Perceptual transformation* from one sensory domain to another: Chinese volume 'sound standards', listening to neural spikes in brain recordings, cloud chambers in radiation studies, visualization of magnetic and electronic recording data.

*C. Multi-modal combination* of sensory domain information: Newton in measuring the speed of sound, Hipparcos satellite using relative 3D stereo imaging.

While these procedures have allowed collaborative measurement (and sharing of knowledge) of object properties they have all occurred within the domain of the human observer $O_{hs}$ and helped to extend the limited sensory capacities. Such a use of the human (or any other) observer as reference carries basic assumptions about the observer itself and the object of observation, and introduces an irrecoverable bias or a priori knowledge to the physical measures, concepts and theories developed. Due to the lack of a general theory of observation these biases have so far remained unknown and they have not been formally described or they have just been neglected. The early history of modern physics includes notions about the perceptual aspects of observation and measurement that are mainly metaphorical in nature and have very little to do with real theories of perception. It seems possible that in quantum mechanics an explicit general theory of the observer could have theoretical value in constructing descriptions of the quantum-level observation processes and the entanglement between the observer and the object of observation.

Summary

The perception-related origins of physical measures and standards are considered within the framework of the general observer theory. The impact of observer characteristics on the development of observer-centric physics, physical concepts and metrics are analyzed. A preliminary theoretical approach is suggested for the construction of a general observer theory and formulation of its relationship to observer-centered physical concepts and theories. The approach makes it possible to construct a theory of the observer that is intrinsic in any theory of physics.

Bibliography

Göte Nyman received his PhD in 1983 from University of Helsinki and worked as professor of general psychology, specialized in human perception. He is the founder the research group POEM (Psychology of Evolving Media and Technology, www.poem-research.org). His current interests include e.g. vision and image quality, quality experience, psychology of the virtual, collaboration, and hci. He is a long-time member of the Finnish Pattern Recognition Society (Hatutus).